\documentclass[preprint,showpacs,preprintnumbers,amsmath,amssymb]{revtex4}

\usepackage{graphicx}      
\usepackage{dcolumn}      
\usepackage{bm}             
\usepackage{amssymb}    

\begin{document}

\title{Monopole-center vortex chains in $SU(2)$ gauge theory}
\author{Seyed~Mohsen~Hosseini Nejad} \altaffiliation {smhosseininejad@ut.ac.ir}
\author{Sedigheh Deldar} \altaffiliation {sdeldar@ut.ac.ir}
\affiliation{Department of Physics, University of Tehran, P.O. Box 14395/547, Tehran 1439955961, Iran}

\begin{abstract}
We study the relation between center vortex fluxes and monopole fluxes for $SU(2)$ gauge group in a model. This model is the same as the thick center vortex model but we use monopole-antimonopole configurations instead of center vortices in the vacuum. Comparing the group factor for the fundamental representation obtained by monopole-antimonopole configurations with the one obtained by the center vortices, we conclude that the flux between the monopole-antimonopole can be split into two center vortex fluxes. Studying the potentials induced by monopole-antimonopole configurations and center vortices, we obtain monopole-vortex chains which appear in lattice Monte Carlo simulations, as well. We show that two similarly oriented center 
vortices inside the monopole-antimonopole configuration repel each other and make a monopole-vortex chain. 
\end{abstract}


\maketitle

\section{\label{sec:level1}Introduction}

The center vortices are color magnetic line-like (surface-like) objects
in three (four) dimensions which are quantized in terms of center elements of the gauge group. Condensation of center vortices in the vacuum of QCD leads to quark confinement such that  
the color electric flux between quark and antiquark is compressed into tubes and a linear
rising potential between static quarks is obtained. In the vortex picture, quark confinement emerges due to the interaction between center vortices and Wilson loops \cite{Greensite2003,Engelhardt2005}. On the other hand,
monopoles are playing the role of agents of confinement in the dual superconductor scenario \cite{Suzuki1993,Hooft1981}. Therefore, one may expect that there are some kind of relations between monopoles and center vortices. Monte Carlo simulations \cite{Del Debbio1998} indicate that a center vortex
configuration after transforming to maximal Abelian gauge and then Abelian projection, appears in the form of the
monopole-vortex chains in $SU(2)$ gauge group. The idea of monopole-vortex chains has been studied by so many researchers \cite{Del Debbio1998,Cornwall1998,Chernodub2009,Chernodub2005,Reinhardt}. 

In this article, monopole-vortex chains in $SU(2)$ gauge group are investigated in a model. This model is the same as the thick center vortex model \cite{Fabe1998}, but we use monopole-antimonopole flux instead of the center vortex flux. The motivation is to see if by this simple model we can observe the idea of monopole-vortex chains which has already been confirmed by lattice calculations, as well as some other phenomenological models. In this model, monopole-antimonopole configurations which are line-like and similar to center vortices are assumed to exist in the vacuum. Studying the group factors of the monopole-antimonopole configurations and center vortices, we understand that the monopole-antimonopole configurations are constructed of two center vortices.
Increasing the thickness of the center vortex core increases the energy of
 the center vortex and therefore the energy of the vacuum which is made of these vortices.  As a result, the potential energy between static quark-antiquark increases. This fact can be confirmed by this model, as well. This is a trivial fact from the physical point of view that the condensation of vortices leads to quark confinement. Classically, it is very similar to the Aharanov-Bohm effect where increasing the thickness of the magnetic flux and therefore the magnetic energy of the system, changes the interference pattern. Using this simple model, we have calculated the potentials induced by the monopole-antimonopole configurations and center vortices.
Comparing these potentials, we observe that the monopole-antimonopole configurations leads to a larger static quark-antiquark potential compared with the case when we use two center vortices in 
 the model. We interpret this extra energy as a repulsive energy between two center vortices constructing the monopole-antimonopole configurations and then we discuss that the monopole-antimonopole configurations can deform to the monopole-vortex chains as confirmed by lattice calculations and other phenomenological models.

 In section \ref{sec:monopole}, the formation of monopoles which is related to the Abelian gauge fixing method is reviewed in $SU(2)$ gauge group. A model with structures of center vortices and monopole-antimonopole configurations are studied in sections \ref{sec:model1} and \ref{sec:model2}. Then in section \ref{sec:SU(2)}, we study the group factors and potentials of these structures to argue monopole-vortex chains. Finally, we summarize the main points of our study in section~\ref{sec:conclusion}.

\section{Abelian gauge fixing and magnetic monopole charges}\label{sec:monopole}

By Abelian gauge fixing, magnetic monopoles are produced in a non Abelian gauge theory. Specific points in the space where the Abelian gauge fixing becomes undetermined are sources of magnetic monopoles. In the following the formation of the magnetic charge by Abelian gauge fixing method is discussed \cite{Ripka}.

 In order to reduce a non Abelian gauge theory into an Abelian gauge theory, the gluon field under a gauge transformation can not be diagonalized. In fact, the gluon field $A^\mu $ has four components and only one of them can be aligned simultaneously. Therefore, a scalar field is used to fix a gauge. 
One can consider a scalar field $\Phi \left( x\right)$ in the adjoint representation of $SU(N)$ as the following: 
\begin{equation}
\Phi \left( x\right) =\Phi _a\left( x\right) T_a  \label{phigauge}
\end{equation}
where $T_a$ are the $N^2-1$ generators of the $SU\left( N\right) $
gauge group. A gauge which diagonalizes the matrix $\Phi \left( x\right)$ is called Abelian gauge.

Now, we consider the $SU(2)$ gauge group. A gauge transformation $\Omega \left( x\right)$ can diagonalize the field $\Phi \left( x\right)$: 
\begin{equation}
\Phi =\Phi _{a}T_{a}\rightarrow \Omega \Phi \Omega ^{\dagger }=\lambda
T_{3}=\left( 
\begin{array}{cc}
\lambda & 0 \\ 
0 & -\lambda
\end{array}
\right),
\end{equation}
where
\begin{equation}
\lambda =\sqrt{\Phi _{1}^{2}+\Phi _{2}^{2}+\Phi _{3}^{2}}.
\end{equation}
The eigenvalues $\lambda \left( x\right) $ of the matrix $\Phi \left(
x\right) $ are degenerated when $\lambda =0$ and therefore three components $%
\Phi _{a=1,2,3}\left( \vec{r}\right) $ are zero at specific points $\vec{r}=\vec{r}_{0}$: 
\begin{equation}
\Phi _{1}\left( \vec{r}_{0}\right) =0\;\;\;\;\;\;\Phi _{2}\left( \vec{r}%
_{0}\right) =0\;\;\;\;\;\;\Phi _{3}\left( \vec{r}_{0}\right) =0
\label{x123r}
\end{equation}
In the vicinity of the point $\vec{r}=\vec{r}_{0}$ we can express $\Phi \left( \vec{%
r}\right)$  in terms of a Taylor expansion:
\begin{equation}
\Phi \left( \vec{r}\right) =\Phi _{a}\left( \vec{r}\right)
T_{a}=T_{a}C_{ab}\left( x_{b}-x_{0b}\right),
\end{equation}
where $C_{ab}=\left. \frac{%
\partial \Phi _{a}}{\partial x_{b}}\right| _{\vec{r}=\vec{r}_{0}}$. Therefore the field $\Phi \left( \vec{r}\right) $
has the hedgehog shape in the vicinity of the point $\vec{r}=\vec{r}_{0}$. One can define another coordinate system where the point $\vec{r}_{0}$ is placed at the origin. In this coordinate system, the field $\Phi
\left( \vec{r}^{\prime }\right) $ has the form: 
\begin{equation}
\Phi \left( \vec{r}^{\prime }\right) =x_{a}^{\prime
}T_{a}\;\;\;\;\;\;x_{a}^{\prime }=C_{ab}\left( x_{b}-x_{0b}\right).
\label{shedge}
\end{equation}
 Dropping the prime on $%
x^{\prime }$ and using the spherical coordinates for the
vector $\vec{r}$, one get to 
\begin{equation}
\Phi \left( \vec{r}\right) =x_{a}T_{a}=\frac{r}{2}\left( 
\begin{array}{cc}
\cos \theta & e^{-i\varphi }\sin \theta \\ 
e^{i\varphi }\sin \theta & -\cos \theta
\end{array}
\right).
\end{equation}
The gauge transformation $\Omega$ which diagonalizes the
hedgehog field $\Phi $ is
\begin{equation}
\Omega \left( \theta ,\varphi \right)=\left( 
\begin{array}{cc}
e^{i\varphi }\cos \frac{\theta }{2} & \sin \frac{\theta }{2} \\ 
-\sin \frac{\theta }{2} & e^{-i\varphi }\cos \frac{\theta }{2}
\end{array}
\right).   \label{omega}
\end{equation}
Therefore the
hedgehog field $\Phi $ is diagonalized as the following:
\begin{equation}
\Omega \Phi \Omega ^{\dagger }=\frac{r}{2}\left( 
\begin{array}{cc}
1 & 0 \\ 
0 & -1
\end{array}
\right) =rT_{3}.
\end{equation}
The gluon field transforms under the same gauge
transformation:
\begin{equation}
\vec{A}=\vec{A}_aT_a\rightarrow \Omega
\left( \vec{A}+\frac 1{ie}\vec{\nabla}\right) \Omega ^{\dagger }.
\label{aomega}
\end{equation}
One can obtain:

\begin{equation}
\frac 1{ie}\Omega \vec{\nabla}\Omega ^{\dagger }=\frac 1e\left( -\vec{e}%
_\theta T_2e^{i\varphi }-\vec{e}_\varphi \frac{1+\cos \theta }{r\sin \theta }%
T_3+\vec{e}_\varphi \frac 1r\left( \cos \varphi T_1-\sin \varphi T_2\right)
\right).
\end{equation}
Thus, the gluon field under the gauge transformation of Eq. (\ref {omega}) can be separated into a regular part $\vec{A}%
^{R} $ and a singular part: 
\begin{equation}
\vec{A}=\vec{A}_{a}T_{a}=\vec{A}_{a}^{R}T_{a}-\frac{1}{e}\vec{n}_{\varphi }%
\frac{1+\cos \theta }{r\sin \theta }T_{3}.  \label{asph}
\end{equation}
The singular part has the form of a gauge field in the vicinity of a magnetic monopole with magnetic charge equal to:
\begin{equation}
g=-\frac{4\pi }{e}T_{3}.  \label{gpie}
\end{equation}

To summarize, we observe that in the vicinity of the points where the eigenvalues of the matrix $\Phi \left(
x\right) $ are degenerate, the singular part of the gluon field in the Abelian gauge behaves like a monopole with
magnetic charge $g=-\frac{4\pi }{e}T_{3}$.

\section{ A model of vacuum structure }
\label{sec:model1}
In this model \cite{Fabe1998}, the Yang Mills vacuum is dominated by center vortices which have a finite thickness (a core). In $SU(N)$ gauge group, there are $N-1$ types of center vortices corresponding to 
the nontrivial center elements of $z_n=e^{\frac{i2\pi n}{N}}$ enumerated by the value $n=1,...,N-1$. The effect of piercing a Wilson loop by a thick center vortex is assumed to be represented by insertion of a group element $G$ in the link product as the following
\begin{equation}
W(C)=Tr \big[U...U\big]\longrightarrow Tr \big[U...G...U\big],
\label{W2}
\end{equation}
where 
\begin{equation}
\label{average}
G(\vec\alpha_C^{n}(x),{S})={S}\;\exp\left[i\vec{\alpha}_C^{n}(x)\cdot\vec{\mathcal{H}}\right]\;{S}^\dagger.
\end{equation}
 The $\{\mathcal{H}_i~i=1,..,N-1\}$ are the Cartan generators, ${S}$ is a random element of $SU(N)$ gauge group and
 angle $\vec{\alpha}_C^{n}(x)$ shows the flux profile which depends on the Wilson loop size and the location $x$ of the center vortex with respect to the Wilson contour $C$.  
The random group orientations associated
with $S$ are uncorrelated, and should be averaged. The averaged contribution of $G$ over orientations in the group manifold specified by $S$ is
\[
\overline{G}(\vec{\alpha}_C^{n}(x))=\int dS \; S \exp\left[i\vec{\alpha}_C^{n}(x)\cdot\vec{\mathcal{H}}\right] {S}^\dagger= 
\]
 \begin{equation}
 \label{group}
=\frac{1}{d_r}Tr\left(\exp\left[i\vec{\alpha}_C^{n}(x)\cdot\vec{\mathcal{H}}\right]\right)\; \mathbf{I}_{d_r}\equiv \mathcal{G}_r(\vec{\alpha}_C^{n}(x))\; \mathbf{I}_{d_r},
\end{equation}
where $\mathcal{G}_r(\vec{\alpha}_C^{n}(x))$ is called the group factor and $\mathbf{I}_{d_r}$ is the $d_r\times d_r$ unit matrix.  
 In $SU(N)$ case, the group factor of the fundamental representation interpolates smoothly from $e^{\frac{i2\pi n}{N}}$, if the core of the center vortex is located completely inside the Wilson loop, to $1$, if the core is completely exterior. 
 The Wilson loop $C$ is assumed as a rectangular $R\times T$ loop in the $x-t$ plane with $T\gg R$ where the left and the right time-like legs of the
 Wilson loop are located at $x=0$ and $x=R$. In other words, the two static charges are located at these points.
 
 A desired ansatz for angle $\vec\alpha_{C}^{n}(x)$ must lead to a well-defined potential $i.e.$ linearity and Casimir scaling at the intermediate distances. Any reasonable ansatz for the angle $\vec\alpha_{C}^{n}(x)$ must satisfy the following conditions:
 \begin{description}
\item{1.} $\vec\alpha_{C}^{n}(x)= 0$ when a center vortex locates far outside the Wilson loop.

\item{2.} $\vec\alpha_{C}^{n}(x)=\vec\alpha_{max}^{n}$ when a center vortex locates deep inside a large Wilson loop. The maximum value of the angle $\vec\alpha_{max}^{n}$ is
obtained from the following maximum flux condition:
\begin{equation}
 \label{max}
\exp(i\vec{\alpha}_{max}^{n}\cdot\vec{\mathcal{H}_r})=\exp(i{\alpha}_{i{(max)}}^{n}{\mathcal{H}_{ir}})=e^{i2k\pi n /N} I,
\end{equation}
where $k$ is the $N$-ality of representation $r$.
\item{3.} $\vec\alpha_{C}^{n}(x)= 0$ as $R\to0$ (small Wilson loop).
\end{description}

An ansatz for the flux profile which would meet these conditions is assumed as the following \cite{Fabe1998}

\begin{equation}
\alpha_{i}^{n}(x)=\frac{\alpha_{i{(max)}}^{n}}{2}[1-\tanh(ay(x)+\frac{b}{R})],
\label{alpha}
\end{equation}
where $n$ indicates the center vortex type, $a , b$ are free parameters of the model, ${\alpha}_{i{(max)}}^{n}$ corresponding to Eq. (\ref {max}) indicates the maximum value of the flux profile and $R$ is the distance between two static charges. $y(x)$, the nearest distance of $x$ from the timelike side of the loop, is 
\begin{equation}
       y(x) = \left\{ \begin{array}{cl}
                     x-R & \mbox{for~} |R-x| \le |x| \cr
                     -x  & \mbox{for~} |R-x| > |x|
                   \end{array} \right.
\end{equation}
The flux of Eq. (\ref{alpha}) is one of the many examples that can give the appropriate potential. Some other examples were discussed in Ref. \cite{Deld2000}.
 
\begin{figure}[h!]
\centering
a)\includegraphics[width=0.49\columnwidth]{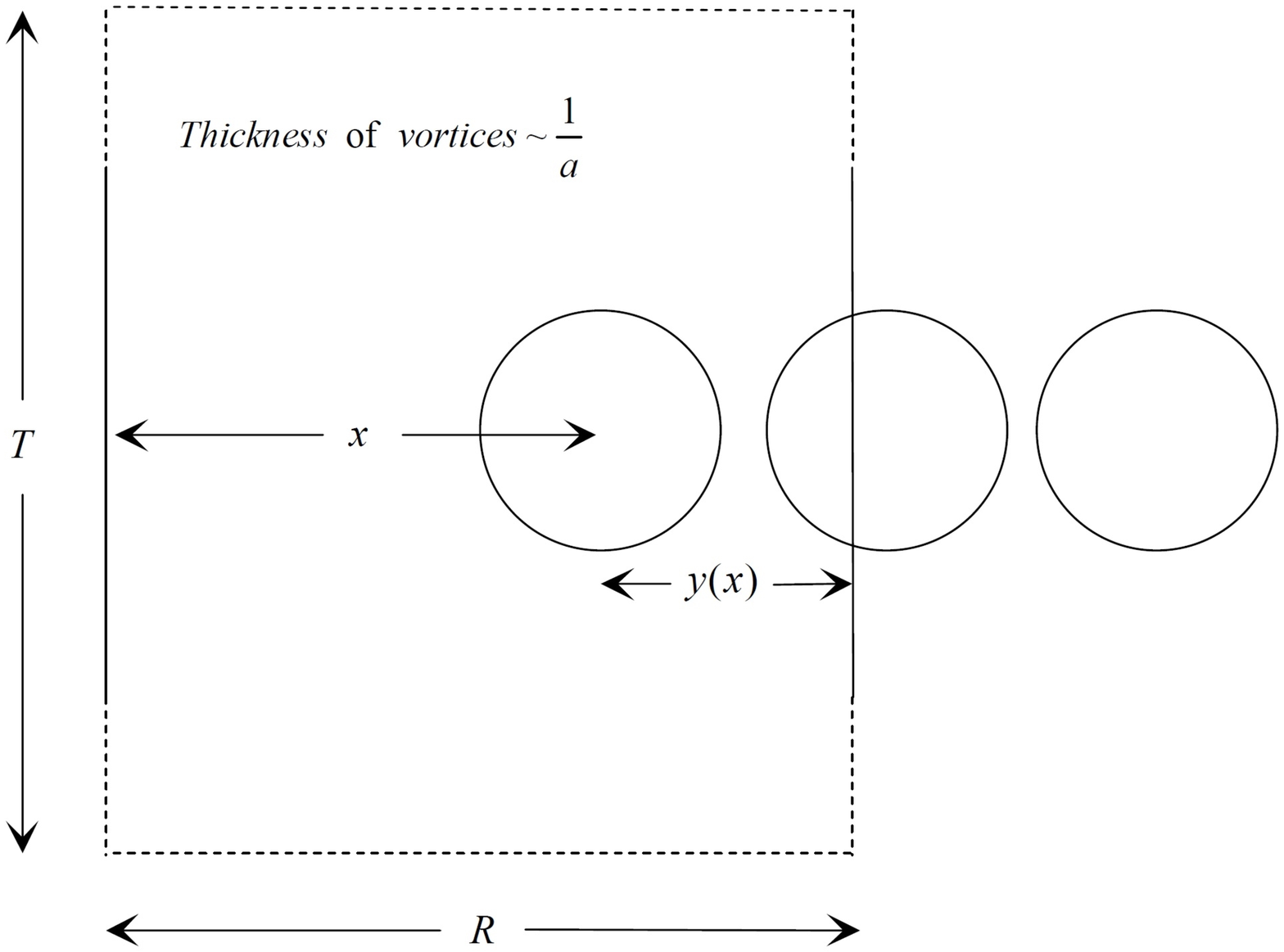}
b)\includegraphics[width=0.46\columnwidth]{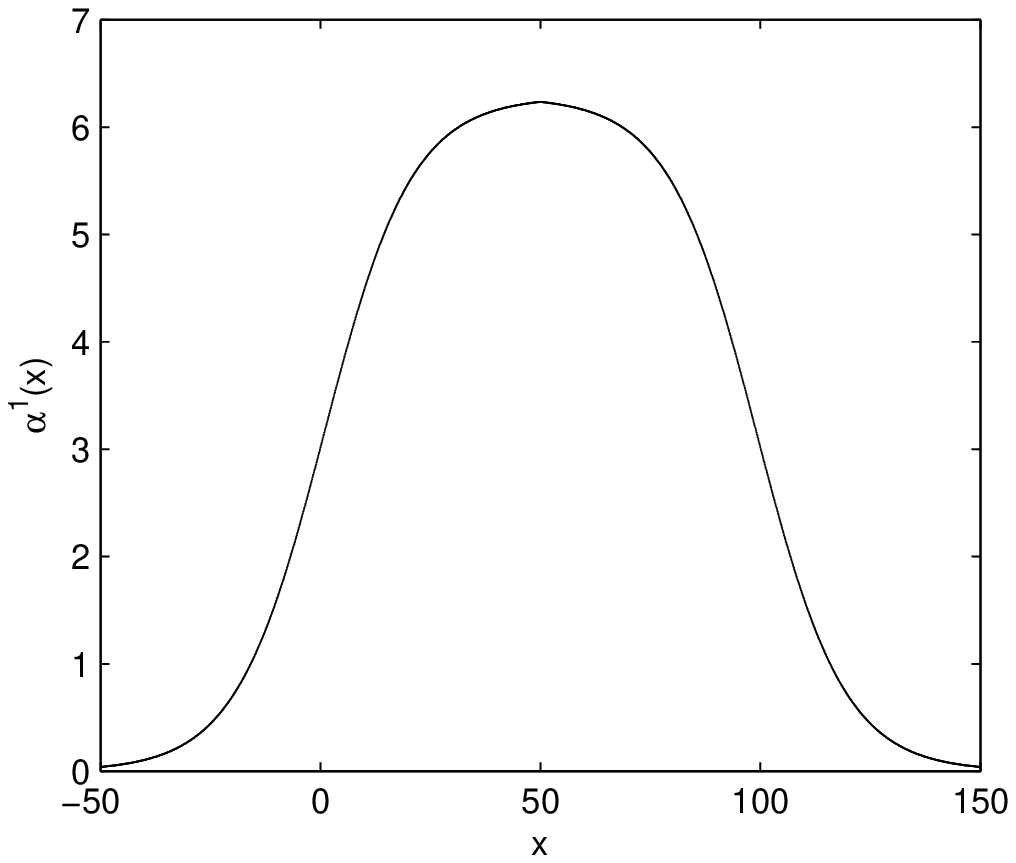}
\caption{a) The figure schematically shows the interaction between the $SU(2)$ center vortices with the ansatz of the flux profile given in Eq. (\ref {alpha3}) and the Wilson loop as a rectangular $R\times T$ loop in the $x-t$ plane as well as some parameters of the ansatz. The effect of center vortex on the loop is assumed by insertion of a group element $G$ in the link product of the Wilson loop which interpolates smoothly between $G=+I$ at $\alpha^1=0$, when the center vortex locates far outside the Wilson loop, and  $G=-I$ at $\alpha^1_{max}=2\pi$, when the center vortex completely locates inside the Wilson loop. b) The angle $\alpha^{1}$ versus $x$ corresponds to the ansatz for $SU(2)$ gauge theory. The left and right time-like legs of the
 Wilson loop are located at $x=0$ and $x=R=100$. The free parameters $a$ and $b$ are chosen to be $0.05$ and $4$, respectively. }\label{0}
\end{figure}

For $SU(2)$ gauge group, when the vortex core is entirely contained within the Wilson loop, using Eq. (\ref {max}), we get
\begin{equation}
\label{alpha2}
\exp[i{\alpha}^{1}_{max} {\mathcal{H}}_{3}]= z_1 I,
\end{equation} 
where $\mathcal{H}_{3}$ is Cartan generator and $z_1 I=e^{\pi i} I$ is the center element of $SU(2)$ gauge group. Therefore, the maximum value of the angle $\alpha^{1}_{max}$ for the fundamental representation is equal to 
$2\pi$. Thus, the ansatz of the flux profile given in Eq. (\ref {max}) for $SU(2)$ is obtained as the following
\begin{equation}
\alpha^{1}(x)=\pi[1-\tanh(ay(x)+\frac{b}{R})].
\label{alpha3}
\end{equation}
 Figure \ref{0}a schematically shows the interaction of center vortices with an $R\times T$ Wilson loop using the ansatz for the flux profile of center vortices which is given in Eq. (\ref {alpha}). The ansatz of the flux profile in Eq. (\ref {alpha3}) is plotted in Fig. \ref{0}b.
In this model, as an assumption, the probabilities of piercing the plaquettes in the Wilson loop by center vortices are uncorrelated. Assuming that an 
$n$th center vortex appears in any given plaquette with the probability $f_n$, the expectation value of the Wilson loop is obtained:
\begin{equation}
\label{w}
<W(C)>=\prod_x \left\{ 1 - \sum_{n=1}^{N-1} f_n
(1- Re\mathcal{G}_r[\vec\alpha^{n}_C(x)]) 
\right\} <W_0(C)>,
\end{equation}

where
\begin{equation}
\label{p}
  f_n = f_{N-n}  ~~~~ \mbox{and} ~~~~ 
       \mathcal{G}_r[\vec\alpha^{n}(x)]) = \mathcal{G}_r^*[\vec\alpha^{N-n}(x)]).                    
\end{equation} 
 $<W_0(C)>$ denotes $Tr \big[U...U\big]$ which no vortex pierces the Wilson loop.
 
 One of the criteria for the color confinement is the area law for the Wilson loop $\it{i.e.}$
 \begin{equation}\label{wilson}
<W(C)>=\exp{\big(-\sigma
A(C)\big)}<W_0(C)>.
\end{equation}
Here $A(C)$ is the minimal surface spanned on the Wilson loop $C$ and
$\sigma> 0$ is the confining string tension. 
Using Eq. (\ref {wilson}) into Eq. (\ref {w}), the string tension is obtained as the following
  \begin{equation}\label{sigma}
 \sigma =- {1\over A} \sum_x \ln\left\{ 1 - \sum_{n=1}^{N-1} f_n
                        (1 - Re\mathcal{G}_r[\vec\alpha^{n}_{C}(x)]) \right\}.
\end{equation}

One gets the static potential induced by center vortices between static color charges in representation $r$ at distance $R$ as the following 
\begin{equation}
\label{potential}
V_r(R) = -\sum_{x=-\infty}^{\infty}\ln\left\{ 1 - \sum^{N-1}_{n=1} f_{n}
(1 - {\mathrm {Re}}\mathcal{G}_{r} [\vec{\alpha}^n_{C}(x)])\right\},
\end{equation} 
where the center of vortex cores pierces the middle of plaquettes $i.e.$ $x=(n+\frac{1}{2})a$ ($n \in (-\infty,\infty)$) where $a$ is the lattice spacing. We use $a=1$ throughout this paper. Although $R$ takes only integer values in the lattice formulation, but the figures related to $V_r (R)$ are platted over the continuous interval.

For $SU(2)$ gauge group, the static potential induced by $z_1$ center vortices at $f_{1} \ll 1$ and small distances between static charges (small $R$) where $\alpha^{1}(x)\ll2\pi$ is obtained as the following

\begin{equation}
\label{potential2}
V_j(R) = \left\{{\frac{f_1}{6}}\sum_{x=-\infty}^{\infty} 
                 {\alpha}^1(x) \right\} j(j+1),
\end{equation}
where spin index $j$ shows the representations in $SU(2)$ gauge theory. According to Eq. (\ref {potential2}), the static potential is proportional to the eigenvalue of the quadratic Casimir operator $i.e.$ $V_j(R) \sim  j(j+1)$ in agreement with the Casimir scaling effect observed in lattice simulations \cite{Ambjorn:1984mb}. The Casimir proportionality of the static potential induced by center vortices can be generalized from $SU(2)$ to $SU(N)$. For observing the property of Casimir scaling in the potentials at intermediate regime, the probability $f_n$ should be far smaller than $1$. Therefore, the probability $f_n$ is chosen $0.1$ as a desired value in the calculations.

The Casimir scaling is not found at intermediate distances for any
choice of the free parameters related to the ansatz in Eq. (\ref {alpha}). But it is observed for a large region of the parameter
space. As an example, the extent of Casimir scaling region at intermediate distances can be changed by any factor $F$ by setting $a \rightarrow a/F,~b \rightarrow bF$. The thickness of the center vortex would be on the order ${1}/{a}$ for the ansatz given in Eq. (\ref {max}). Therefore, choosing $F>1$ as an integer value, increases the thickness of the center vortex and the Casimir scaling region while $F<1$, decreases these quantities.

In the next section, we investigate the effect of a monopole flux on a Wilson loop.
\section{ Monopole-antimonopole configurations }

\label{sec:model2}
\begin{figure}
\begin{center}
\resizebox{0.1\textwidth}{!}{
\includegraphics{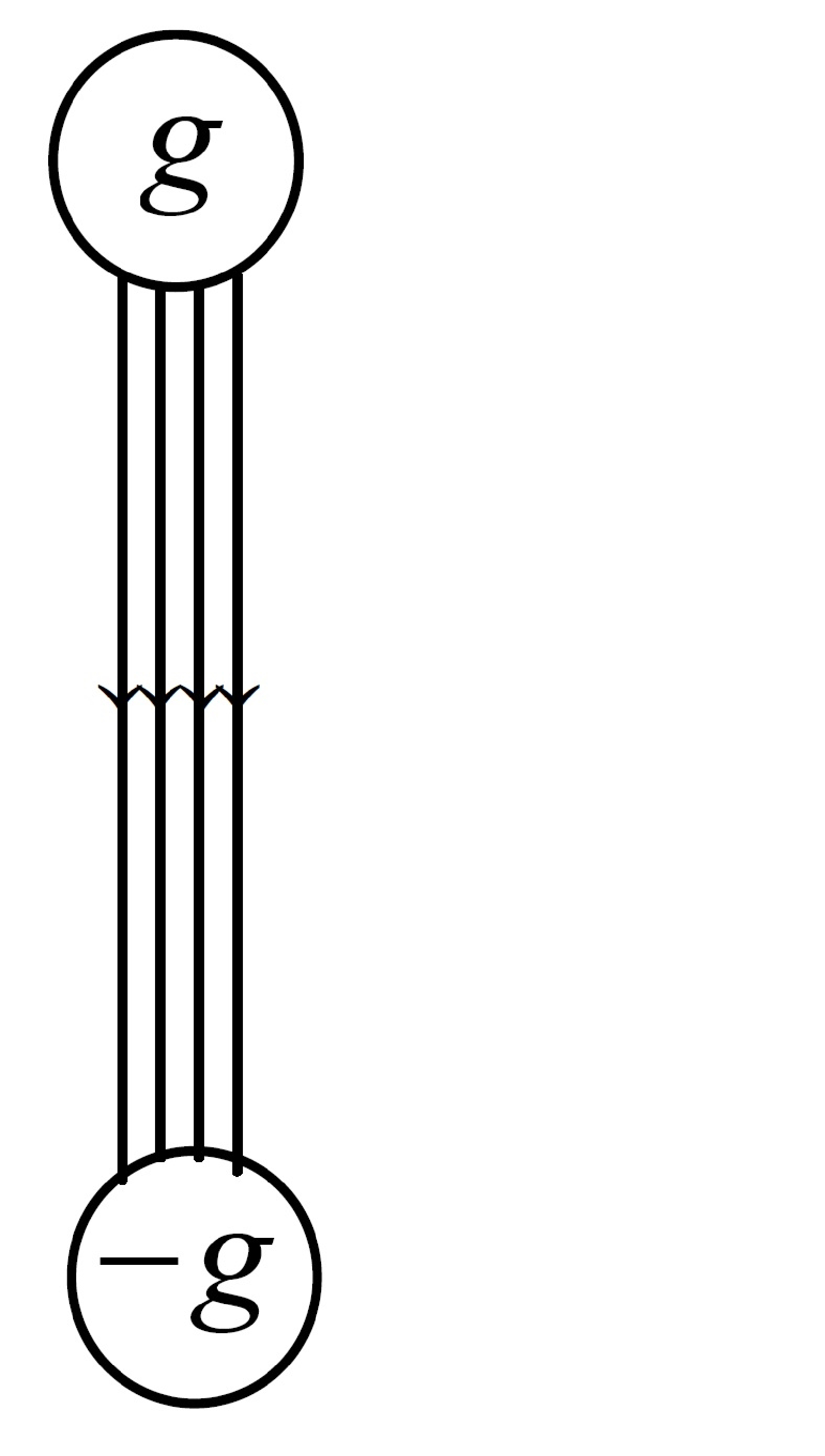}}
\caption{\label{1}
A schematic view of the monopole-antimonopole configuration which is initially considered to be localized. This configuration is line-like, the same as the center vortices. The arrows on the lines show the direction of the magnetic field.}
\end{center}
\end{figure}

 Now, we consider monopole-antimonopole pairs as the Abelian configurations in the vacuum. We are assuming that the magnetic fields between monopole and antimonopole are initially localized in a tube as plotted in Fig. \ref{1}. We use these configurations in the thick center vortex model instead of center vortices. The monopole-antimonopole configurations are line-like similar to the center vortices. The effect of piercing a Wilson loop by the monopole-antimonopole configuration is represented by insertion of a phase $e^{ ie\int_S\vec{B}.d\vec{s}}$ \cite{Chernodub2005} in the link product where $e$ is the color electric charge and $\int_S\vec{B}.d\vec{s}$ is the total magnetic flux of the monopole. The magnetic field of a monopole with topological charge $g$ obeys the Maxwell equation $\vec{%
\nabla}.\vec{B}=g\delta \left( \vec{r}\right) $. Therefore, the total magnetic flux of a monopole crossing the surface $S$  is equal the magnetic charge $g$ as the following \cite{Chatterjeea2014,Ripka}
\begin{equation}
\label{ce}
{\Phi}_m=\int_S\vec{B}.d\vec{s}=\int_Vd^3r\;\vec{\nabla}.\vec{B}=\int_Vd^3r\;g\delta
\left( \vec{r}\right) =g,
\end{equation}
where $V$ is the volume enclosed by the surface $S$. For $SU(2)$ gauge group, $g$ is the monopole charge in Eq. (\ref{gpie}). If we attribute a thickness for these configurations as what is done for the thick center vortices, the effect of a monopole-antimonopole configuration on a Wilson loop is to multiply the loop 
by a group element the same as the one in Eq. (\ref{group}).
If a monopole-antimonopole configuration is entirely contained within the loop, then
\begin{equation}
\label{center}
\exp\left[i\vec{\alpha}^{n}\cdot\vec{\mathcal{H}}\right]=e^{ ieg},
\end{equation}
where $eg$ satisfies the charge quantization condition $eg=2n\pi$. 

For $SU(2)$ gauge group, corresponding to Eq. (\ref {gpie}), the magnetic charge of the monopole is $g=-\frac{4\pi }{e}\mathcal{H}_{3}$ and the one for antimonopole is $g=+\frac{4\pi }{e}\mathcal{H}_{3}$ where $\mathcal{H}_{3}=diag\big(\frac{1}{2},-\frac{1}{2}\big)$ represents the Cartan generator. When a monopole-antimonopole pair is entirely contained within the Wilson loop, using $SU(2)$ magnetic charge into Eq. (\ref {max}), we get
\begin{equation}
\label{alpha4}
\exp[i{\alpha}^{0}_{max} {\mathcal{H}}_{3}]= e^{\pm i4\pi \mathcal{H}_{3}}=e^{i2\pi} I,
\end{equation} 
where index $n=0$ is related to the monopole-antimonopole configurations. The sign in the exponent is not important since the direction of the configuration which pierces the Wilson loop is not important.
Therefore the maximum value of the angle $\alpha^{0}_{max}$ for the fundamental representation is equal to 
$4\pi$. Therefore the ansatz of the flux profile given in Eq. (\ref {max}) for the monopole-antimonopole configurations of $SU(2)$ gauge theory is obtained as the following
\begin{equation}
\alpha^{0}(x)=2\pi[1-\tanh(ay(x)+\frac{b}{R})].
\label{alpha5}
\end{equation}
The potential induced by monopole-antimonopole configurations is the same as the one induced by center vortices represented in Eq. (\ref{potential}). For $SU(2)$ gauge group, the potential induced by monopole-antimonopole configurations for the fundamental representation is obtained as the following
\begin{equation}
\label{potential3}
V_f(R) = - \sum_{x=-\infty}^\infty \ln\{(1-f_0) + 
                      f_0\mathcal{G}_{f}[\alpha^0(x)] \}.
\end{equation}

 In the next section, we study the group factors and the potentials for the center vortices and the monopole-antimonopole configurations.

\section{SU(2) and vacuum structures}
\label{sec:SU(2)}
To study the center vortices and monopole-antimonopole configurations in the vacuum for $SU(2)$ gauge group, we discuss the interaction between the Wilson loop and these configurations. First, the group factors of these configurations which have an important role in producing the potentials of Eq. (\ref{potential}) \cite{Rafibakhsh2014,HD2015} are studied and the relation between these configurations is discussed. Then, with calculating the potentials induced by these configurations, interactions inside these configurations are studied.     

\subsection{Interaction between the Wilson loop and center vortices}

First, we calculate the group factor of the center vortices in $SU(2)$ gauge group. The group factor for the fundamental  representation of 
$SU(2)$ is obtained from Eq. (\ref{group}) 
 \begin{equation}
\mathcal{G}_{j=1/2}={1\over 2j+1}\mbox{Tr}\exp[i{\alpha}^{1} {\mathcal{H}}_{3}]=cos(\frac{{\alpha}^{1}}{2}), 
 \label{group-factor}
 \end{equation}
where ${\mathcal{H}}_{3}$ is the Cartan generator of $SU(2)$ gauge group. According to Eq. (\ref{alpha2}), the maximum value of the angle $\alpha^{1}_{max}$ for the fundamental representation is equal to 
$2\pi$. Using ansatz given in Eq. (\ref{alpha3}), Fig. \ref{23}a shows $\mathcal{G}_r(\alpha^{n})$ obtained from center vortices versus $x$ for a fundamental representation Wilson loop with $R=80$. The legs of  the Wilson 
loop are located at $x = 0$ and $x = 80$. The free parameters $a$ and $b$ are chosen to be $0.05$ and $4$, respectively. The group 
factor interpolates smoothly from $-1$, when the vortex core is located completely inside the Wilson loop, to $1$, when the core is entirely outside the loop. Figure \ref{g12}a shows $\mathcal{G}_r(\alpha^{n})$ obtained from center vortices versus $x$ for small sizes of Wilson loops (small $R$). For small size of the Wilson loop, center vortices are partially located inside the Wilson loop and the maximum flux is not center vortex flux. Therefore the minimum of the group factor of center vortices increases with decreasing the size of the Wilson loop and approaches to $1$.
\subsection{Interaction between the Wilson loop and monopole fluxes}

\begin{figure}[h!]
\centering
a)\includegraphics[width=0.46\columnwidth]{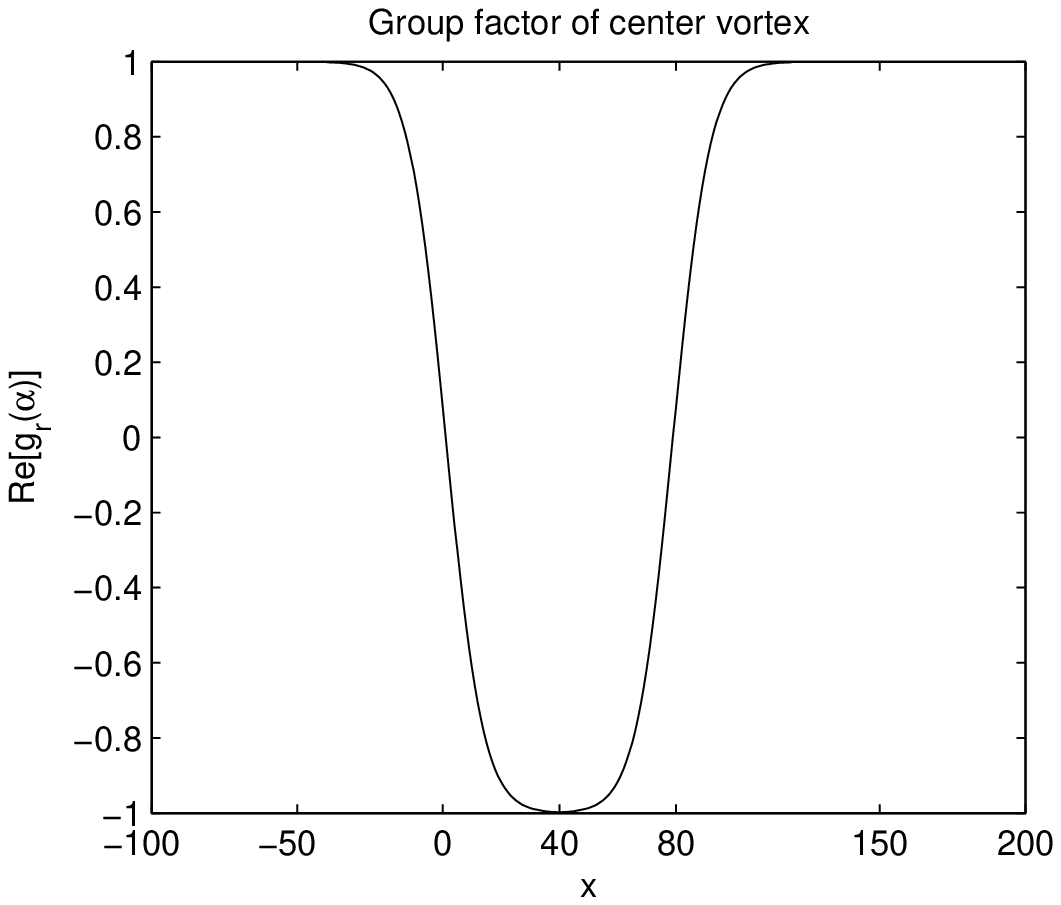}
b)\includegraphics[width=0.46\columnwidth]{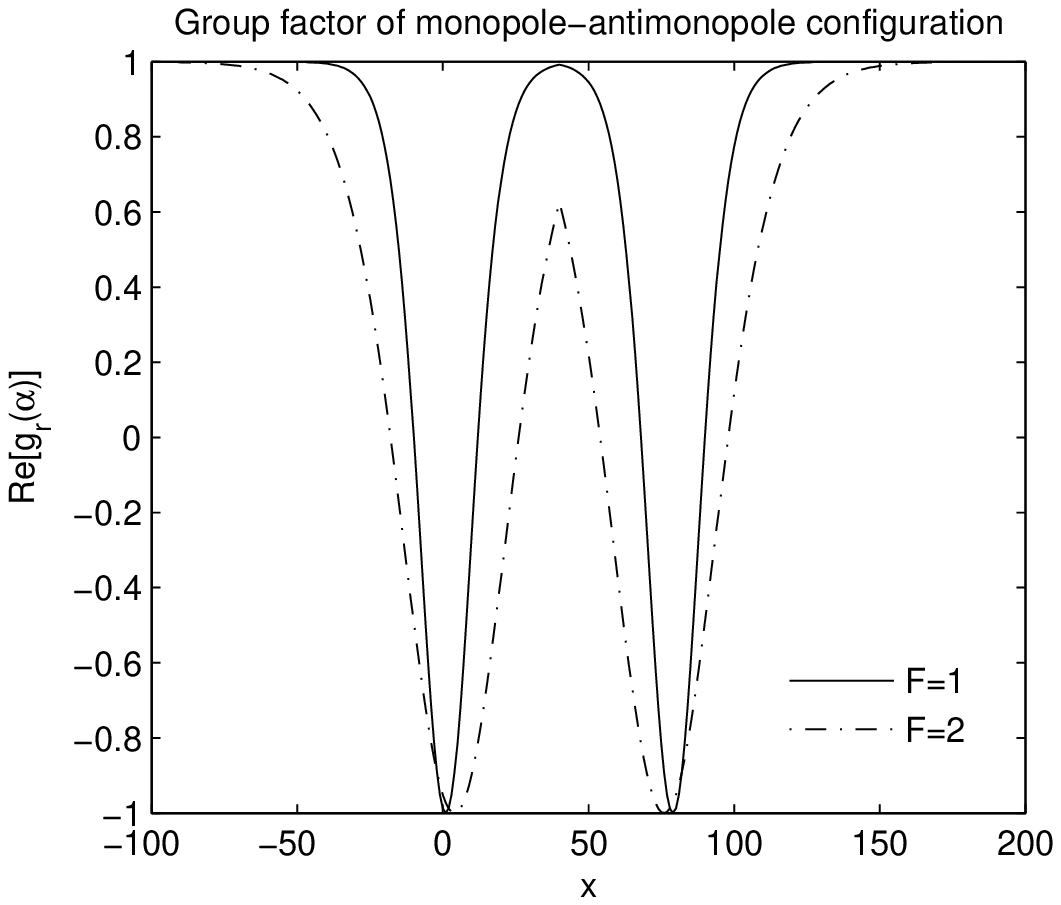}
\caption{ a) ${\mathrm {Re}}(\mathcal{G}_{r})$ obtained from center vortices versus $x$ for the fundamental representation of the $SU(2)$ gauge group for $R=80$. The free parameters $a$ and $b$ are chosen to be $0.05$ and $4$, respectively. When center vortices locate completely inside the Wilson loop, the value of the group factor is $-1$  b) the same as a) but obtained from the monopole-antimonopole configurations. The free parameters $a$ and $b$ are chosen to be $ 0.05/F$ and $ 4F$, respectively. When half of the core of the monopole-antimonopole configuration locates inside the Wilson loop (at $x = 0$ or $x = 80$), the flux inside the loop is equivalent to the center vortex flux. Therefore the fluxes of the center vortices inside the monopole-antimonopole configuration do not have an overlap. It seems that two similarly oriented
 vortices repel each other. As shown, by changing the free parameters (for example, varying $F=1$ to $F=2$), the results are the same and half of the core of the monopole-antimonopole configuration is equivalent to the center vortex flux. By varying $F=1$ to $F=2$, the thickness of center vortices is increased and becomes more than the size of the Wilson loop ($R=80$).  Therefore, when the center of vortex core is located in the middle of the Wilson loop ($x=40$), $\mathcal{G}_r(\alpha^{0})$ becomes less than $1$. Increasing the size of the Wilson loop, the maximum value of the group factor reaches to $1$. We show in Fig. \ref{4} that by changing $F=1$ to $F=2$ the static potentials are just scaled up.
  }\label{23}
\end{figure}

Next, we calculate the group factor of the monopole-antimonopole configurations in $SU(2)$ gauge group. Using Eq. (\ref{group}), the group factor for the fundamental representation is obtained as the following
  \begin{equation}
\mathcal{G}_{f}=cos(\frac{{\alpha}^{0}}{2}), 
 \label{gfactor}
 \end{equation}
According to Eq. (\ref{alpha4}), the maximum value of the angle $\alpha^{0}_{max}$ for the fundamental representation is equal to 
$4\pi$. Using ansatz given in Eq. (\ref{alpha5}), Fig. \ref{23}b plots $\mathcal{G}_r(\alpha^{n})$ obtained from the monopole-antimonopole configurations versus $x$ for a Wilson loop of the size $R=80$ for the fundamental representation. The Wilson 
loop legs are located at $x = 0$ and $x = 80$. The free parameters $a$ and $b$ are chosen to be $0.05/F$ and $ 4F$, respectively.  When the monopole-antimonopole configuration overlaps the minimal area of the Wilson loop, it affects the loop. For $F=1$, the value of the group 
factor is $1$, when the monopole-antimonopole configuration core is located completely inside or outside the Wilson loop. For $F>1$, the thickness of center vortices is increased and the maximum value of the flux profile $\alpha^{0}$ is less than $4\pi$. Therefore, when the center of vortex core is located in the middle of the Wilson loop with the size $R=80$, $\mathcal{G}_r(\alpha^{0})$ becomes less than $1$. Increasing the size of the Wilson loop, the maximum value of the group factor reaches to $1$. 

\begin{figure}[h!]
\centering
a)\includegraphics[width=0.46\columnwidth]{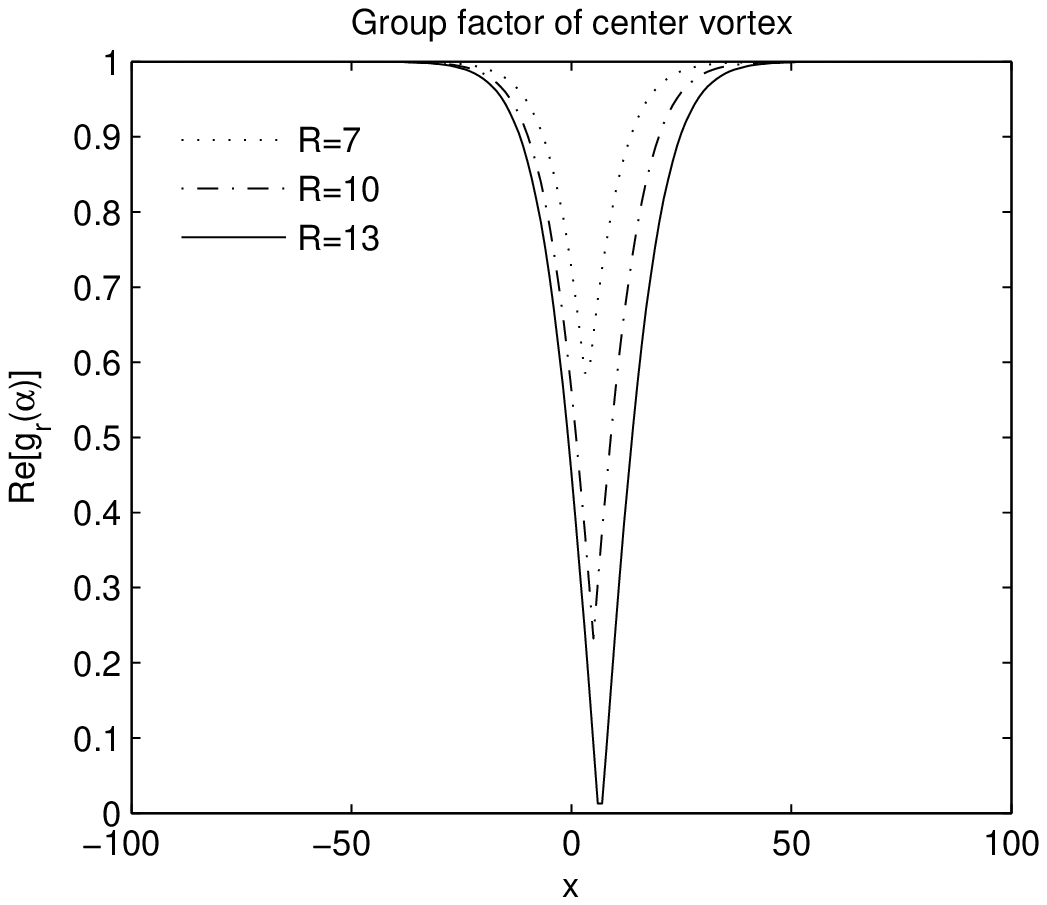}
b)\includegraphics[width=0.46\columnwidth]{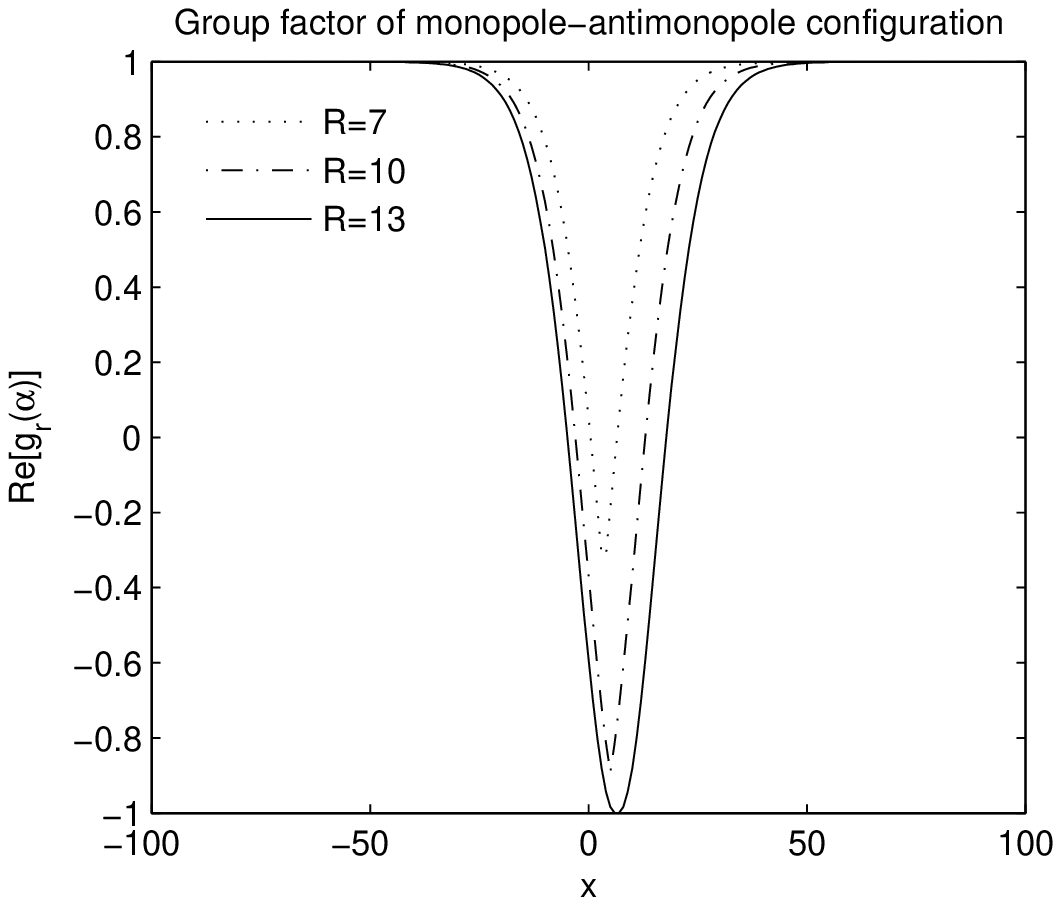}
\caption{a) ${\mathrm {Re}}(\mathcal{G}_{r})$ obtained from center vortices versus $x$ for the fundamental representation of the $SU(2)$ gauge group for small sizes of Wilson loops (small $R$). The free parameters $a$ and $b$ are chosen to be $0.05$ and $4$, respectively. Decreasing the size of the Wilson loop, the minimum value of the group factor increases and approaches to $1$. b) the same as a) but obtained from the monopole-antimonopole configurations. Assuming the monopole-antimonopole configuration is constructed from center vortices, we observe that the value of $-1$ for the group factor, corresponding to the total flux of the center vortex, happens when $R$ is equal to $13$. Decreasing the size of the Wilson loop, the minimum of the group factor of the monopole-antimonopole configuration increases and deviates from  $-1$.}\label{g12}
\end{figure}

When the center of monopole-antimonopole configuration is placed
at $x = 0$ or $x = 80$, half of the maximum flux enters the Wilson loop. The group 
factor interpolates smoothly from $1$, when the monopole-antimonopole configuration core is entirely outside the loop, to
$-1$, when the half of the core is located inside the Wilson loop. As shown in Fig. \ref{23}b, the results are the same by changing the free parameters. This behavior of the group factor is similar to the group factor of the center vortex which changes smoothly between $1$, when the center vortex is located completely outside the Wilson loop and $-1$, when the center vortex is completely inside the loop. Therefore, half of the monopole-antimonopole flux is equal to the vortex flux $\it{i.e.}$ the monopole-antimonopole configuration is constructed from two center vortices. 
Figure \ref{g12}b shows $\mathcal{G}_r(\alpha^{n})$ obtained from monopole-antimonopole configuration versus $x$ for small sizes of Wilson loops. For small $R$, monopole-antimonopole configurations are partially located inside the Wilson loop and the maximum flux is not equal to the total flux of the monopole-antimonopole configuration. Assuming the monopole-antimonopole configuration is constructed from center vortices, we observe that the value of $-1$ for the group factor, corresponding to the total flux of the center vortex, happens when $R$ is equal to $13$. Decreasing the size of the Wilson loop, the minimum of the group factor of the monopole-antimonopole configuration increases and deviates from  $-1$.

The assumed center vortices inside the monopole-antimonopole configuration affect each other. Since the behavior of one half of the monopole flux on the Wilson loop is the same as one center vortex, the fluxes of two vortices constructing monopole-antimonopole configuration do not have an overlap. Therefore it seems that two center vortices inside the monopole-antimonopole configuration repel each other. In the next subsection the interaction between these center vortices is investigated, in details. Before that we study another approach, explained in ref. \cite{Deldar2014}, for obtaining the relation between center vortex and monopole fluxes.

  Using fractional flux of a monopole, the flux of a center vortex is constructed in $SU(2)$ gauge group. Substituting $\mathcal{H}_3$ from Eq. (\ref{gpie}) in Eq. ({\ref{alpha2}) and ${\alpha}^{1}_{max}=2\pi$, we get \cite{Deldar2014}
\begin{equation}
\label{cent}
\exp\left[i2\pi \mathcal{H}_3\right]=\exp\left[-ie\frac{g}{2}\right]=z_1 I.
\end{equation}
According to Eq. ({\ref{ce}), $g$ is equal to the total magnetic flux of a monopole. Therefore the effect of a center vortex on the Wilson loop is the same as  the effect of an Abelian configuration corresponding to the half of the matrix flux $g$ on the Wilson loop.

Now, we obtain the flux of this Abelian configuration and compare it with the flux of center vortex which is equal to ${\Phi}_v=\pi$ \cite{Ambjorn2000}.

 The contribution of this Abelian configuration on the Wilson loop is
   \begin{equation}
\label{c8}
W=\mathcal{G}_f=\frac{1}{d_f}\mathrm{Tr}\left(\exp\left[-ie\frac{g}{2}\right]\right)=\frac{1}{2}\mathrm{Tr}\left(\begin{array}{cc} e^{-i\pi} & 0 \\0 &e^{i\pi} 
\end{array} \right)=e^{i\pi}.
\end{equation}  
Comparing Eq. ({\ref{c8}) with the contribution
of an Abelian field configuration to the Wilson loop which is $W=e^{iq{\Phi}}$ ($q$ means units of the electric charge and  $q=1$ for the fundamental representation) \cite{Chernodub2005}, the flux of this Abelian configuration is equal to $\pi$.

Therefore, the flux of this Abelian configuration corresponding to the half of the magnetic charge $g$, is the same as one center vortex on the Wilson loop. 

 In the next subsection the interaction between center vortices inside the monopole-antimonopole configuration is studied.

\subsection{Monopole-vortex chains}

 In the previous sections, we have shown that the flux between a monopole-antimonopole pair is constructed from the fluxes of two vortices. To understand the interaction between two center vortices inside the monopole-antimonopole configuration, we study the potentials induced by the center vortices and the monopole-antimonopole configurations using the ``center vortex model". Using Eqs. ({\ref{potential}) and ({\ref{potential3}), Fig. \ref{4} shows the static potential of the fundamental representation at intermediate distances induced by monopole-antimonopole configurations compared with the one induced by the center vortices. The potential energy induced by monopole-antimonopole configurations is larger than the twice of the potential energy induced by the center vortices. The free parameters $a$, $b$ and $f_n (n=0,1)$ are chosen to be $ 0.05/F$, $ 4F$ and $0.1$, respectively. As shown in Fig. \ref{4}, the result do not change by varying the factor $F$ related to the free 
 parameters. Using two center vortices in the model without any interaction, potential at small distances is obtained to be equal to the case when we use one center vortex with a thickness of twice the original one. We recall that increasing the thickness of the center vortex core would increase the energy of
 the center vortex and therefore the energy of the vacuum which is made of these vortices.  As a result, the potential energy between static quark-antiquark increases. This is shown in figure \ref{4}.  On the other hand, if there is no interaction between the vortices of the monopole-antimonopole pair, the induced potential for the small distances is expected to be equal to the induced 
 potential by the two non interacting vortices. However, as shown in figure \ref{4}, the induced potentials are not equal. The extra energy obtained for the induced potential between the quark-antiquark using monopole-antimonopole configurations, can be interpreted as a repulsion energy between the two center vortices constructing the configuration. Therefore, two 
vortices with the same flux orientations inside the monopole-antimonopole configuration repel each other. 

The interaction between the constructing vortices of monopole-antimonopole pair can be observed by the small or intermediate size Wilson loops. For large enough Wilson loops, two center vortices, constructing the monopole-antimonopole configuration, are located completely 
inside the Wilson loop. Thus, the effect of two center vortices ($z^2$) on the large Wilson loops is trivial ($z^2=I$). The flat potential at large distances in Fig. \ref{chain}, shows this trivial behavior.  Therefore, interaction between two vortices can not be observed for $R$ greater than the vortex core size. The vortex core size is about $20$ with the free parameters we used in the model. We recall that the lengths are dimensionless in the model. Thus, interaction between vortices of the monopole-antimonopole configuration is possible for $R$ less than $20$.    
 
Using the monopole-antimonopole configuration of 
the vacuum, we only show that there is a repulsion between two center vortices
 within the monopole-antimonopole configuration and they construct a monopole-vortex chain. 
However, these monopole-vortex chains should be observed in $3$ dimensions. In the model, since 
the Wilson loop is a rectangular $R\times T$ loop in the $x-t$ plane, it probably intersects  
with one of the legs of the chain at a time. Many random piercings of the Wilson loop by these legs 
and then averaging those random piercings leads to the confinement. In fact, the repulsion deforms the localized flux and only one of the vortices would intersect the Wilson loop as confirmed by the chain models \cite{Del Debbio1998,Cornwall1998,Chernodub2009,Chernodub2005,Reinhardt}. These cases are shown in Fig. \ref{chain}. 
In ref. \cite{Reinhardt}, Reinhardt and $\it{et~ al.}$ explained that the monopole-antimonopole flux splits into two equal portions of center vortex fluxes shown in Fig. \ref{Dirac}. The Wilson loop which intersects one of these center vortices leads to confinement for the static sources. 

In addition, the dual superconductor picture of quark confinement was proposed by Nambu in  1970's \cite{Nambu1974}. Ginzburg-Landau theory defines two parameters: the superconducting coherence length $\displaystyle \xi$ and the London magnetic field penetration depth $\displaystyle \lambda$. 

As an interesting possibility, the repel of two center vortices may mean
the Type-II superconductor of the QCD vacuum, that is, the Ginzburg-Landau
parameter $\kappa=\lambda /\xi$ of the QCD vacuum is larger than $1/\sqrt{2}$.                                                                                      

We would like to mention that the interaction between vortices has been
studied by the domain model (the modified thick center vortex model) in ref. \cite{HD2015}, as well. In that article, based on ``energetics'' 
we have shown that two vortices with the same flux orientations inside $(z_1)^2$ vacuum domains repel each other.  While two vortices with opposite flux orientations inside $z_1z^*_1$ vacuum domains attract each other. The group factors analysis of $(z_1)^2$ and $z_1z^*_1$ vacuum domains agree with this article. Since
two similarly oriented vortices inside $(z_1)^2$ vacuum domain repel each other, we conclude that they
do not make a stable configuration and one should consider each of them as a single vortex
in the model. On the other hand, since
two vortices with the opposite orientation inside $z_1z^*_1$ vacuum domain attract each other, we conclude that they
 make a stable configuration. Adding the contribution of the $z_1z^*_1$
 vacuum domain to the potential obtained
from center vortices, the length of the Casimir scaling regime increases \cite{HD2015}. The results of this paper is in agreement with our previous paper.

To summarize, in this article, we obtain a chain of monopole-vortex. The magnetic flux coming from a monopole inside the  chain is 
squeezed into vortices of finite thickness and a non-orientable closed loop is formed. The non-orientable closed loop means that two vortex lines inside the loop have different orientations of magnetic fluxes. Figure \ref{5} schematically shows the interaction between center vortices inside the monopole-antimonopole configuration. 

 \begin{figure}[h!]
\centering
a)\includegraphics[width=0.46\columnwidth]{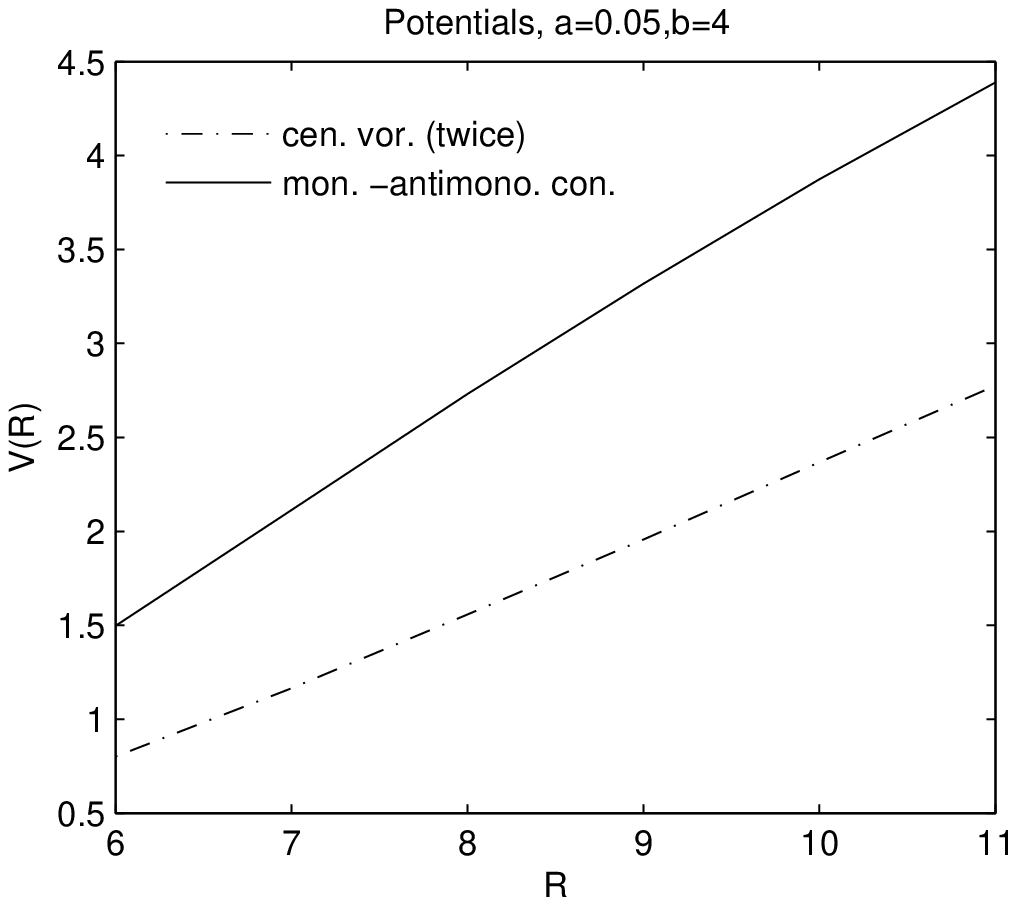}
b)\includegraphics[width=0.46\columnwidth]{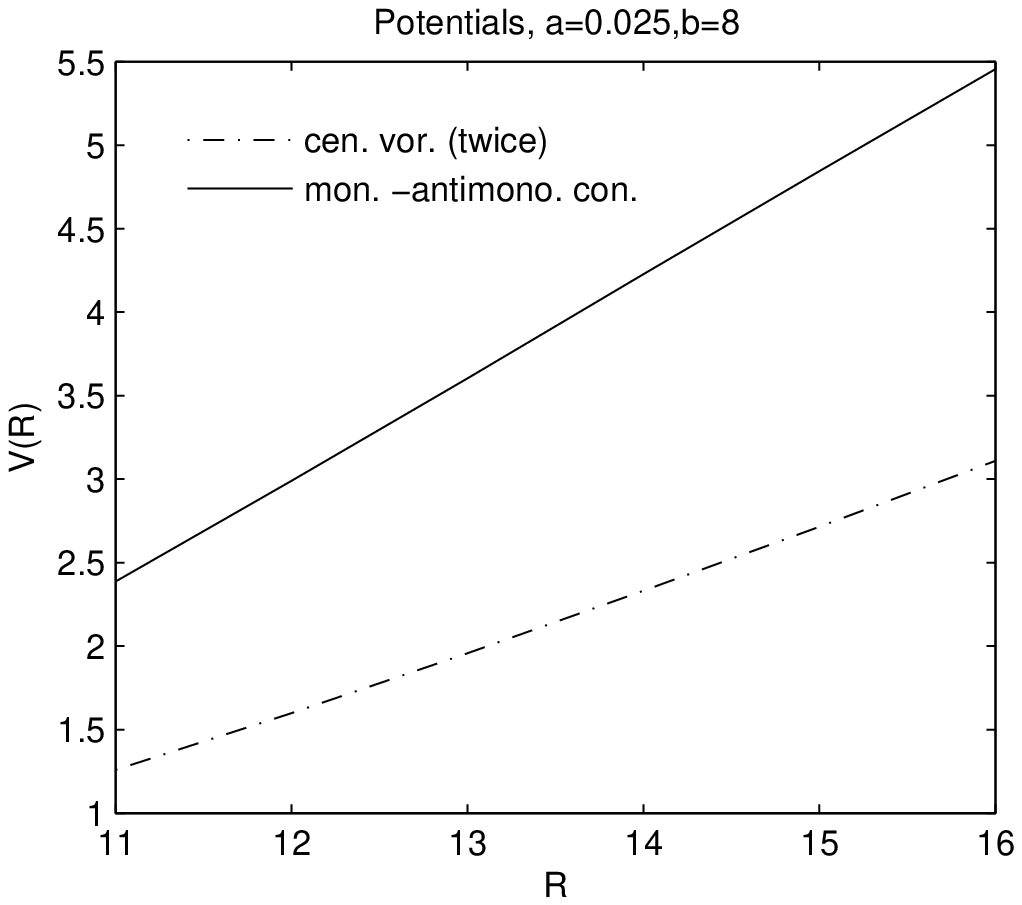}
\caption{a) The potential energy $V_{mon}(R)$ induced by monopole-antimonopole configurations and the twice of the potential $V_{vor}(R)$ obtained by the center vortices. The free parameters $a$ and $b$ are chosen to be $ 0.05/F$ and $4F$ where $F=1$ and the probability $f_n$ ($n=0,1$) is chosen $0.1$. The potential ratio $V_{mon}(R)/2V_{vor}(R)$ is about $1.5
$. The extra positive potential 
energy of static potential induced by monopole-antimonopole configurations compared with the twice of the static potential obtained from center vortices shows that two similarly oriented center 
vortices inside the monopole-antimonopole configuration repel each other and make a monopole-vortex chain. b) the same as a) but for $F=2$. The potential ratios $V_{mon}(R)/2V_{vor}(R)$, obtained from a) and b), are the same within the errors. Therefore, varying the free parameters do not change the physical results.}\label{4}
\end{figure}  

\begin{figure}
\begin{center}
\resizebox{0.46\textwidth}{!}{
\includegraphics{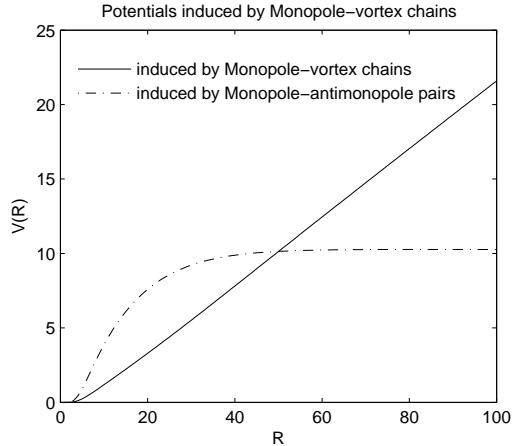}}
\caption{\label{chain}
The potential energy induced by the monopole-vortex chains. The free parameters $a$, $b$ and $f_n$ are chosen to be $ 0.05$, $4$ and $0.1$. If the monopole-vortex chains intersect in two points with the large Wilson loop, the static sources are screened at large distances. On the other hand, if one leg of the monopole-vortex chain intersects the large Wilson loop, confinement is observed for the fundamental representation. }
\end{center}
\end{figure}
\begin{figure}[h!]
\centering
a)\includegraphics[width=0.26\columnwidth]{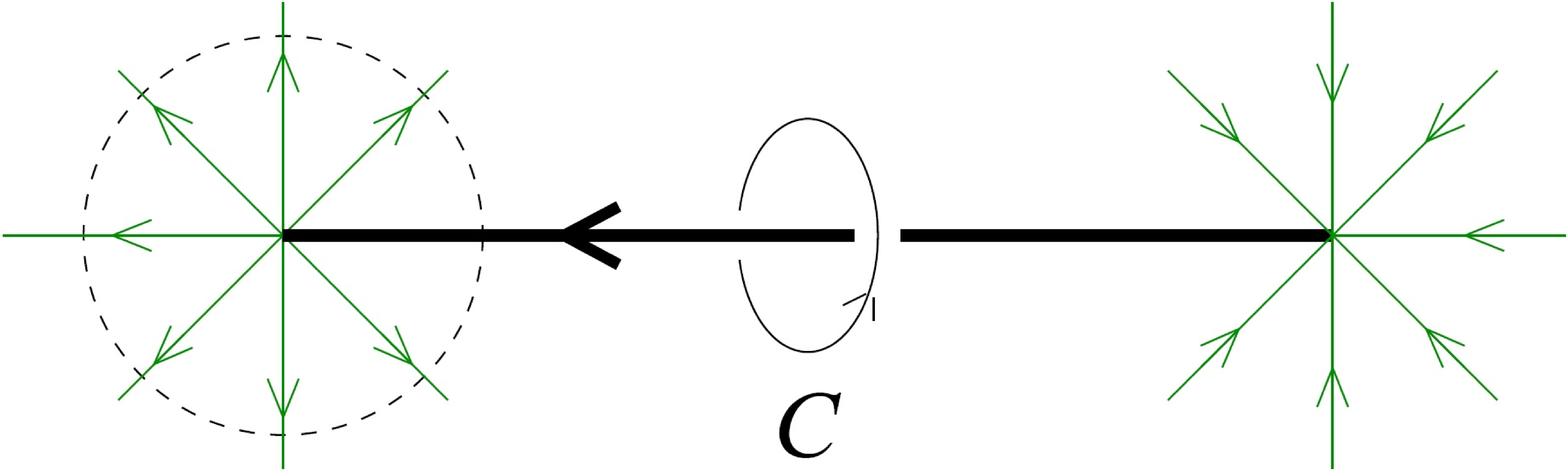}
b)\includegraphics[width=0.26\columnwidth]{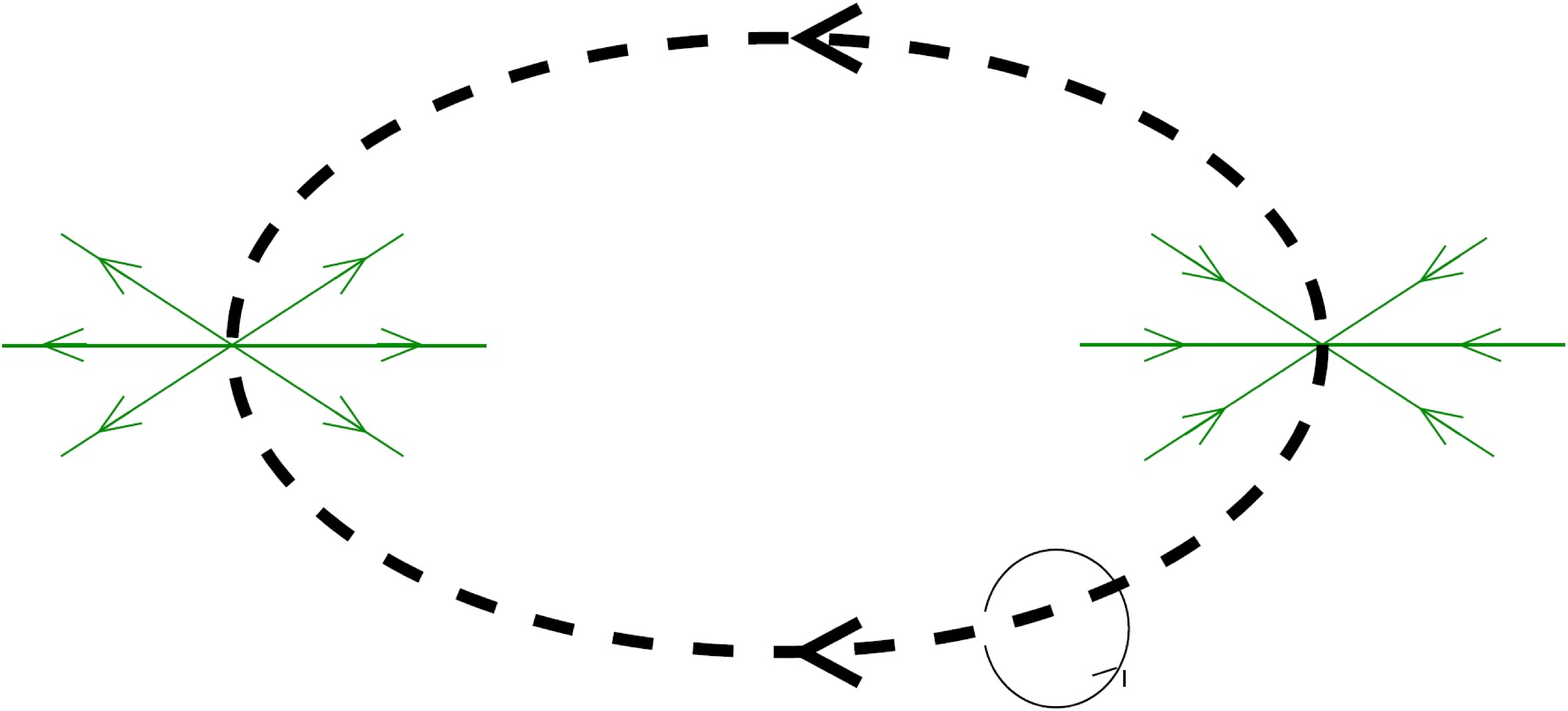}
\caption{ a) The monopole-antimonopole pair in $SU(2)$ gauge group. These configurations contribute unity to the  Wilson  loop $C$. b) Assuming that the magnetic flux of the monopole-antimonopole configuration is split into two center vortex fluxes, one leg of this chain contributes $-1$ to the Wilson loop $C$. Therefore these chains lead to the confinement for the static sources \cite{Reinhardt}. }\label{Dirac}
\end{figure} 

\begin{figure}
\begin{center}
\resizebox{0.05\textwidth}{!}{
\includegraphics{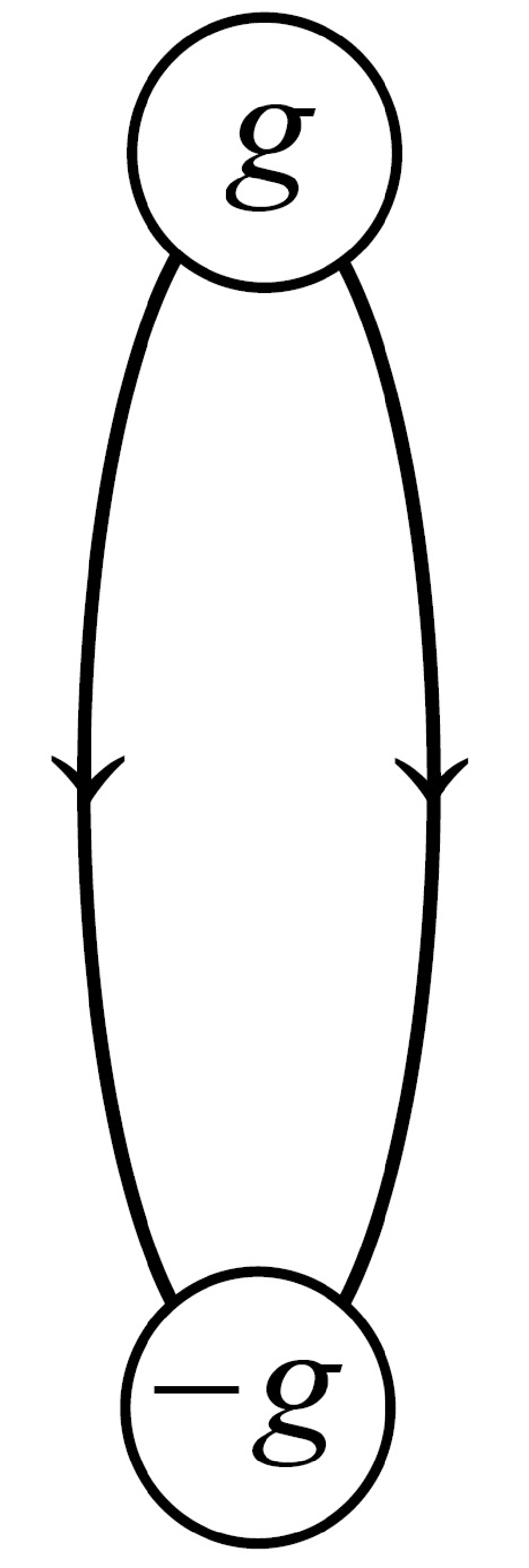}}
\caption{\label{5}
A schematic view of a monopole-vortex chain obtained from the monopole-antimonopole configuration. Center vortices of the monopole-antimonopole configuration repel each other and make a monopole-vortex chain. The arrows on the vortex lines show the direction of the magnetic field of the vortex.}
\end{center}
\end{figure}

Our understanding of monopole-antimonopole flux and monopole vortex chain is also in agreement with other research about this topic as comes in the following.
According to the Monte Carlo simulations, after Abelian projection almost all monopoles are sitting on top of the vortices \cite{Del Debbio1998,Ambjorn2000} as shown in Fig. \ref{6}. Therefore a center vortex upon Abelian projection would appear in the form of monopole-vortex chains. Indeed Abelian monopoles and center vortices correlate with each other.  Figure \ref{7}a shows some monopole-vortex chains in $SU(2)$ gauge group \cite{Ambjorn2000}.
\begin{figure}
\begin{center}
\resizebox{0.5\textwidth}{!}{
\includegraphics{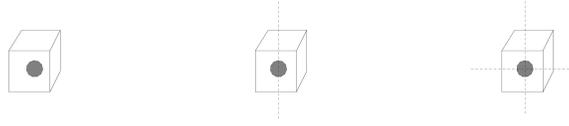}}
\caption{\label{6}
monopoles priced by P-vortices. Almost all monopoles (about 93\%) are priced by one P-vortex (middle panel). Only very small fractions of monopoles either are not pierced 
   at all (about 3\%)(left panel), or are pierced by more than one line (about 4\%)(right panel) \cite{Del Debbio1998}.}
\end{center}
\end{figure} 
\begin{figure}[h!]
\centering
a)\includegraphics[width=0.36\columnwidth]{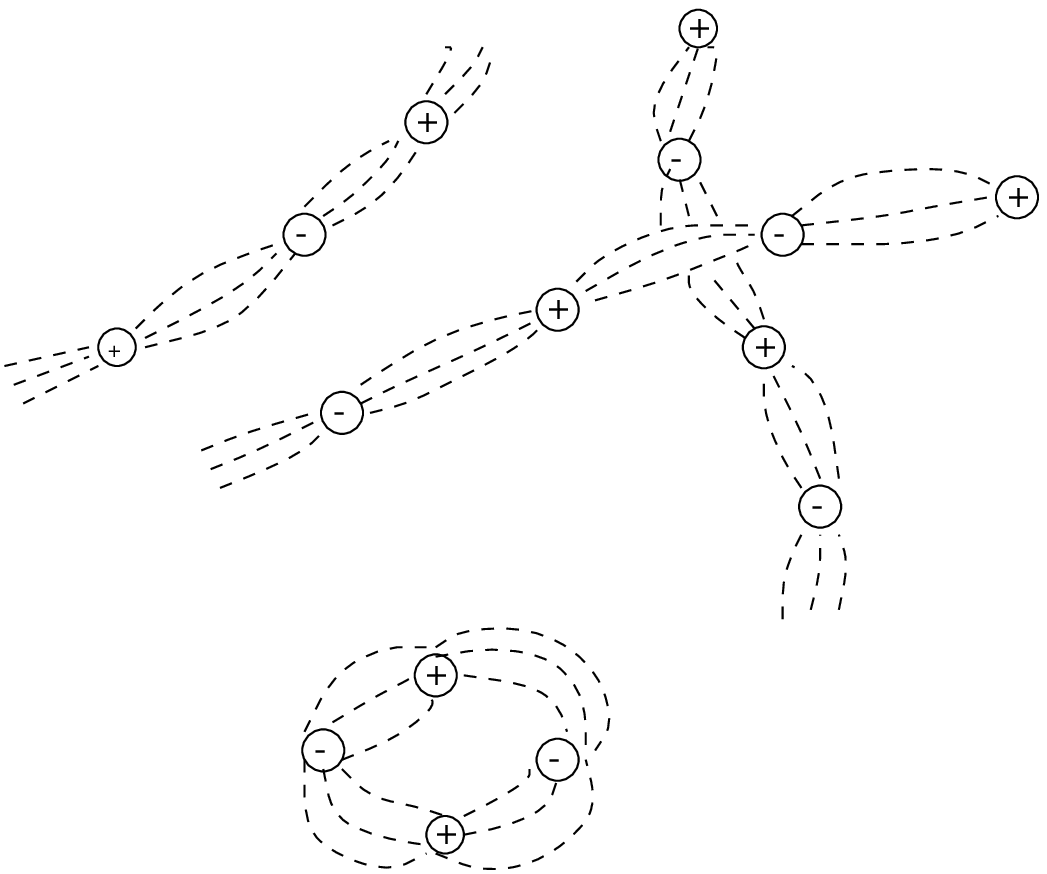}
b)\includegraphics[width=0.1\columnwidth]{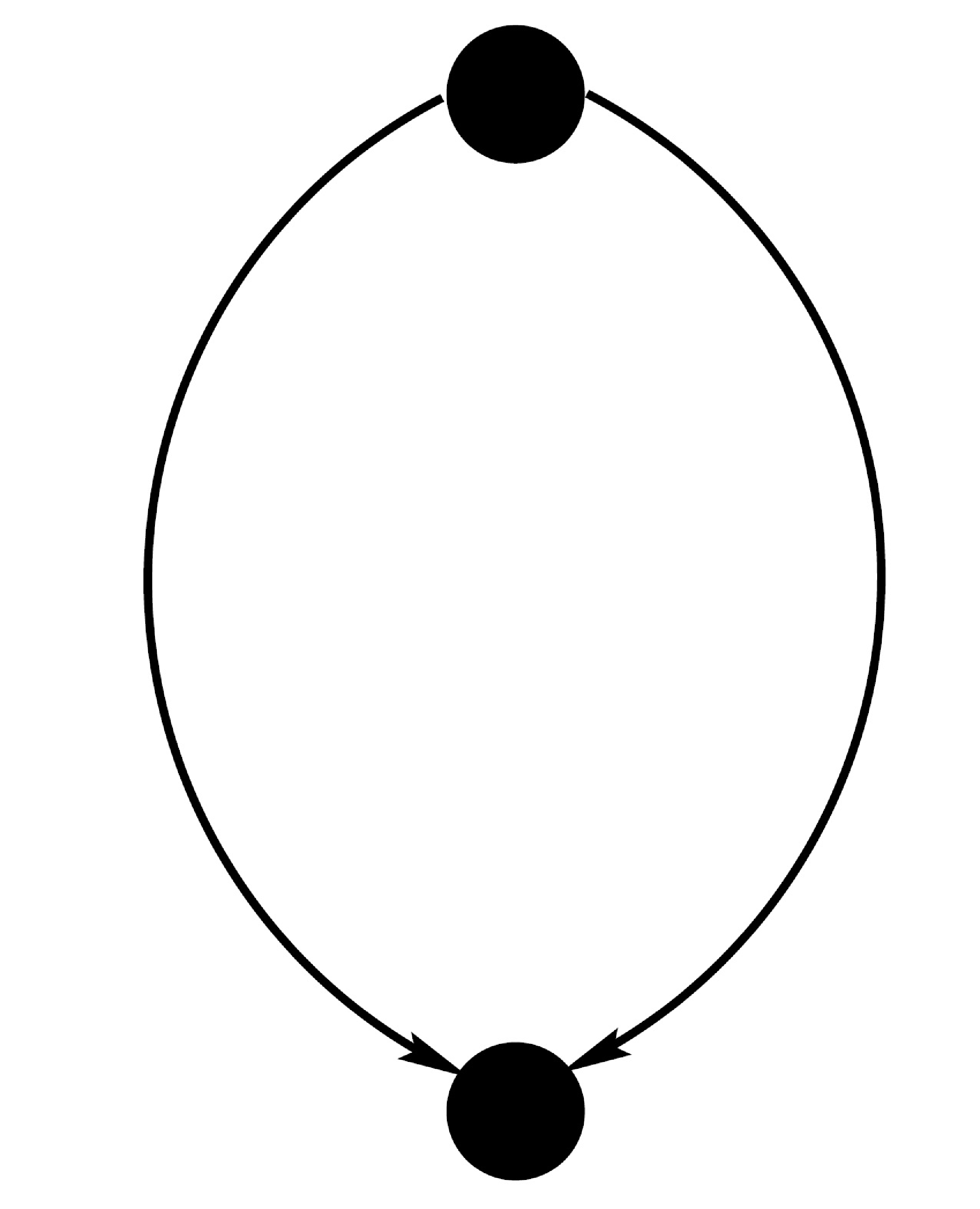}
\caption{a) Some monopole-vortex  chains in $SU(2)$ gauge group shown in ref. \cite{Ambjorn2000}. b) The monopole-vortex  chain shown in ref. \cite{Cornwall1998}. Therefore, the monopole-vortex chain obtained in this article agrees with the results of lattice gauge theory and chain models.}\label{7}
\end{figure}
In addition, the monopole-vortex junctions called as nexuses are studied in ref. \cite{Cornwall1977}. Some solutions to the equations of motion obtained from the low-energy effective energy functional 
$E$ of QCD \cite{Cornwall1998} are studied. Several thick vortices meet at a monopole-like center (nexus), 
with finite action and non-singular field strengths. In $SU(N)$ gauge group each nexus is the source of $N$ center vortices.
Figure \ref{7}b shows monopole-vortex chain obtained by Cornwall for $SU(2)$ gauge group \cite{Cornwall1998}. In ref. \cite{Chernodub2009}, examples of the monopole-vortex chains are also plotted using the method of ref. \cite{Cornwall1998}.

Therefore the monopole-vortex chain in the vacuum obtained from the model agrees with the results of lattice gauge theory and chain models.
\section{conclusion}
 \label{sec:conclusion}

The formation of the monopole-vortex chains which are observed in lattice simulation is studied in a model. This model is similar to the thick center vortex model but instead of center vortices in the model we use monopole-antimonopole configurations which are line-like the same as center vortices. Comparing group factors of monopole-antimonopole configurations and center vortices, we observe that the flux of the monopole-antimonopole configuration is constructed from two center vortex fluxes. Calculating the induced quark-antiquark potential from two non interacting vortices and
monopole-antimonopole configurations and comparing the plots, we observe that the potential energy induced by the monopole-antimonopole configurations is larger than the twice of the one induced by the two non interacting center vortices configurations. The extra positive energy is interpreted as the repulsive energy between the vortices inside the monopole-antimonopole configuration.
The resulting monopole-vortex chains agree with the lattice calculations and phenomenological models. 
In general, these monopole-vortex chains should be observed in $3$ dimensions. In the model,  
the Wilson loop, which is a rectangular $R\times T$ loop in the $x-t$ plane, probably intersects 
with one of the legs of the chain at a time. Many random piercings of the Wilson loop by these legs 
and then averaging those random piercings leads to the confinement.

\section{\boldmath Acknowledgments}

We are grateful to the Iran National Science Foundation (INSF) and the research council of the University of Tehran for
supporting this study.

\end{document}